\documentclass{ws-procs9x6-cpt25}
\begin{document}
	
	\newcommand{\refeq}[1]{(\ref{#1})}
	\def\etal {{\it et al.}}
	
	\title{Nuclear $\mu-e$ conversion via Lorentz and CPT violation}
	
	\author{William P.\ McNulty}
	
	\address{Department of Physics, Indiana University, Bloomington, IN 47408, USA}
	
	\begin{abstract}
		One of the three common experimental channels searching for charged lepton flavor violation (CLFV) via $\mu-e$ conversion involves first capturing the muon on a nuclear target. This channel provides a unique opportunity to constrain four-point quark-lepton operators that may contribute to CLFV, in addition to the electromagnetic operators that can also be accessed in the other channels. We investigate the leading Lorentz- and CPT-violating contributions for nuclear $\mu-e$ conversion experiments within the Standard-Model Extension and obtain the first bounds on the relevant quark-lepton operators using the results of the SINDRUM II experiment; we also consider possibilities for improved constraints from upcoming experiments.
	\end{abstract}
	
	\bodymatter
	
	\section{Background}
	Lepton flavor conservation is an accidental symmetry of the Standard Model (SM), arising as a consequence of leaving the neutrinos massless. The existence of neutrino oscillations due to the nonvanishing masses of at least two neutrino species manifestly violates lepton flavor conservation, and implies that charged lepton flavor violation (CLFV) should also occur. However, without extending the SM beyond the inclusion of massive neutrinos, such processes are expected to be highly suppressed by the smallness of the neutrino masses and the GIM mechanism, with branching ratios $\lesssim 10^{-54}$.\cite{br} Since this is well beyond current sensitivities, any observation of CLFV would be interpreted as a signal of new physics beyond the SM; thus, such processes have been the subject of an intense program of experimental searches.
	
	In the muon-electron sector of CLFV searches, one of the so-called ``golden channels'' is the coherent conversion of a muon to an electron in the presence of a nucleus, $\mu^- +~ ^A_ZN\rightarrow e^- +~ ^A_ZN$; here, ``coherent'' means that the nucleus remains in its ground state throughout the interaction. The experiment is performed by colliding a muon beam with an atomic target, on which some of the muons will capture to form short-lived muonic atoms, with the muon quickly falling to the 1S ground state. The majority of these muons will be captured by the nucleus through the exchange of a $W$ boson: $\mu^- +~ ^A_ZN \rightarrow \nu_\mu +~ ^A_{Z-1}N$. The remainder are expected to decay in orbit, $\mu^- \rightarrow e^- \overline{\nu_e} \nu_\mu$, with the resulting electron exhibiting the energy spectrum corresponding to a three-body decay. The goal of the experiment, however, is to observe the CLFV process $\mu^- +~ ^A_ZN\rightarrow e^- +~ ^A_ZN$. This produces a monoenergetic electron with energy $E^{\mathrm{conv}}_e = m_\mu -E_{\mathrm{bind}}-E_{\mathrm{recoil}}$, where $m_\mu$ is the muon mass, $E_{\mathrm{bind}}$ is the binding energy of the muonic atom in the 1S state, and $E_{\mathrm{recoil}}$ is the (negligible) nuclear recoil energy.
	
	To date, of course, this process is not known to have been observed; therefore, experiments set upper bounds on the ratio of the rate of the CLFV process to the capture rate:
	\begin{equation}
		R_{\mu e} \equiv \frac{\omega_{\mathrm{conv}}}{\omega_{\mathrm{capt}}} = \frac{\Gamma(\mu^- +~ ^A_ZN\rightarrow e^- +~ ^A_ZN)}{\Gamma(\mu^- +~ ^A_ZN \rightarrow \nu_\mu +~ ^A_{Z-1}N)}.
		\label{eq:Rue}
	\end{equation}
	The current best limit on this ratio has been set by the SINDRUM II experiment at the Paul Scherrer Institute (PSI): $R_{\mu e} < 7\times 10^{-13}$ at $90\%$ confidence level (CL) on a gold ($^{197}_{79}\mathrm{Au}$) target.\cite{sind} The next generation of experiments consists of COMET at the Japan Proton Accelerator Complex (J-PARC) and Mu2e at the Fermi National Accelerator Laboratory (Fermilab), both on aluminum ($^{27}_{13}\mathrm{Al}$) targets. COMET will target a sensitivity of $R_{\mu e} \simeq 7\times 10^{-15}$ in Phase I, with a final goal of $R_{\mu e} \simeq 2.6\times 10^{-17}$, both at $90\%$ CL.\cite{com} Mu2e expects a sensitivity of $R_{\mu e} \simeq 6.2\times 10^{-16}$ from Run I, with a final goal of $R_{\mu e} \simeq 8\times 10^{-17}$, both at $90\%$ CL.\cite{mu2e}
	
	\section{Conversion Rate in the Standard-Model Extension}
	The Standard-Model Extension (SME) is the effective field theory that expands on the SM by incorporating potential violations of Lorentz invariance and the CPT symmetry. The leading contributions to nuclear $\mu - e$ conversion in the SME are expected to come from electromagnetic operators of mass dimension five and four-point quark-lepton operators of mass dimension six. At tree level, the quark-lepton operators are uniquely accessed in this experimental channel, while the electromagnetic operators can also be constrained by searches in the other ``golden channels.''
	
	To account for the presence of the nucleus, one first needs to obtain the leptonic wavefunction, which can be written
	\begin{equation}
		\psi_{\kappa,s}(r,\theta,\phi) =
		\begin{pmatrix}
			g(r)\chi_{\kappa,s}(\theta,\phi) \\
			if(r)\chi_{-\kappa,s}(\theta,\phi)
		\end{pmatrix}.
	\end{equation}
	The eigenspinors satisfy $(\boldsymbol{\sigma}\cdot\mathbf{l}+I)\chi_{\kappa,s} = -\kappa \chi_{\kappa,s}$ and $J_z \chi_{\kappa,s} = s\chi_{\kappa,s}$ for $\kappa = \mp(J+1/2)$. The radial wavefunctions are obtained by numerically solving the Dirac equation using spherically-symmetric nuclear charge distributions.\cite{chrg-dist,nucl}
	
	The conversion rate for coherent conversion via the electromagnetic operators can be written as
	\begin{equation}
		\omega_{\mathrm{conv}}^{(5)} = \frac{1}{2} \sum_{s=\pm\frac{1}{2}} \sum_{\kappa = \pm 1} \sum_{s^\prime=\pm\frac{1}{2}} \left| \int \,d^3x ~\bar{\psi}_{\kappa,s^\prime}^{(e)} \mathcal{O}^{\alpha\beta} \psi_{s}^{(\mu)} F_{\alpha\beta}  \right|^2,
		\label{eq:d5rate}
	\end{equation}
	where $F_{\alpha\beta}$ is the electromagnetic field strength tensor containing the electric field of the nucleus, $\bar{\psi}_{\kappa s^\prime}^{(e)}$ is the continuum wavefunction of an electron of energy $E^{\mathrm{conv}}_e$ and spin $s^\prime$, and $\psi_{s}^{(\mu)}$ is the 1S bound-state wavefunction of a muon of spin $s$. The SME operators are denoted by $\mathcal{O}^{\alpha\beta} \in \Bigl\{ (m^{(5)}_{F})^{\alpha\beta}_{\mu e} ,i(m^{(5)}_{5F})^{\alpha\beta}_{\mu e}\gamma^5 ,(a^{(5)}_{F})^{\mu\alpha\beta}_{\mu e}\gamma_{\mu} ,(b^{(5)}_{F})^{\mu\alpha\beta}_{\mu e}\gamma^5\gamma_{\mu} ,\frac{1}{2}(H^{(5)}_{F})^{\mu\nu\alpha\beta}_{\mu e} \sigma_{\mu\nu} \Bigr\}$, with the restriction that one index is temporal and the other spatial (as only the electric field is considered).
	
	The conversion rate via the quark-lepton operators can be written as
	\begin{equation}
		\omega_{\mathrm{conv}}^{(6)} = \frac{1}{2} \sum_{s=\pm\frac{1}{2}} \sum_{\kappa = \pm 1} \sum_{s^\prime=\pm\frac{1}{2}} \left| \int \,d^3x \left( \alpha\bar{\psi}_{\kappa,s^\prime}^{(e)}\mathcal{K}\psi_{s}^{(\mu)}+\beta\bar{\psi}_{\kappa,s^\prime}^{(e)}\mathcal{K}_0\psi_{s}^{(\mu)} \right) \right|^2,
		\label{eq:d6rate}
	\end{equation}
	where the nuclear matrix elements that are nonvanishing for coherent conversion\cite{nucl} are denoted by $\alpha = \sum_q \left< N|\bar{\psi}^{(q)}\psi^{(q)}|N \right>$ for $q \in \{u,d,s\}$ and $\beta = \sum_q \left< N|\bar{\psi}^{(q)}\gamma_0 \psi^{(q)}|N \right>$ for $q \in \{u,d\}$. The SME operators are denoted by $\mathcal{K} \in \Bigl\{ (k^{(6)}_{SV})^{\lambda}_{qq\mu e}\gamma_\lambda ,(k^{(6)}_{SA})^{\lambda}_{qq\mu e}\gamma^5\gamma_{\lambda} ,\frac{1}{2}(k^{(6)}_{ST})^{\kappa\lambda}_{qq\mu e} \sigma_{\kappa\lambda} \Bigr\}$ and $\mathcal{K}_0 \in \Bigl\{ (k^{(6)}_{VS})^{t}_{qq\mu e} ,i(k^{(6)}_{VP})^{t}_{qq\mu e}\gamma^5 ,\frac{1}{2}(k^{(6)}_{VV})^{t\lambda}_{qq\mu e}\gamma_{\lambda} ,(k^{(6)}_{VA})^{t\lambda}_{qq\mu e}\gamma^5\gamma_{\lambda} ,\frac{1}{2}(k^{(6)}_{VT})^{t\kappa\lambda}_{qq\mu e} \sigma_{\kappa\lambda} \Bigr\}$.
	
	Although initial calculations may be done in the lab frame in spherical coordinates, ultimately the results must be transformed to the Sun-Centered Frame (SCF). The SCF is defined such that $T = 0$ corresponds to the 2000 vernal equinox, the $X$ axis points from the Earth to the Sun at that time, and the $Z$ axis is parallel to the rotation axis of the Earth.\cite{SCF-1} The transformation to a standard laboratory frame ($x$ axis towards local south, $y$ axis towards local east) on Earth is given in Eq.\ (7) of Ref.\ \refcite{KPS} and in general depends on the colatitude of the laboratory, the sidereal frequency of the Earth, and the sidereal time (shifted from $T$ by an amount that depends on longitude, among other effects).\cite{SCF-2} The experiments considered here are taken to have $2\pi$ azimuthal coverage. Acceptance in the polar angle is restricted to a region symmetric about $\theta = \pi/2$ with respect to the beamline, which defines the polar $z$ axis; this is related to the $z$ axis in the standard lab frame by an additional improper rotation.
	
	\section{Results and Outlook}
	We obtain constraints on the relevant SME operators using the results of the SINDRUM II experiment. A summary of these results is provided in Table \ref{tab}; detailed results, along with projections for the initial phases of COMET and Mu2e, are available in Ref.\ \refcite{orig}. In general, we expect to see at least an order of magnitude improvement due to the initial phases of the upcoming experiments, while their final sensitivity goals suggest they could ultimately provide two orders of magnitude improvement.
	
	\begin{table}
		\tbl{Constraints based on SINDRUM II results.}
		{\begin{tabular}{@{}cc@{}}\toprule
				Operator Class & Constraint \\
				\colrule
				Electromagnetic & $<(6\mbox{--}8)\times 10^{-12}\,\mathrm{GeV}^{-1}$ \\
				Scalar Quark-Lepton ($u,d$) & $<(6\mbox{--}7)\times 10^{-13}\,\mathrm{GeV}^{-2}$ \\
				Scalar Quark-Lepton ($s$) & $<(10\mbox{--}15)\times 10^{-13}\,\mathrm{GeV}^{-2}$ \\
				$\gamma_0$ Quark-Lepton ($u,d$) & $<(20\mbox{--}30)\times 10^{-13}\,\mathrm{GeV}^{-2}$ \\\botrule
		\end{tabular}}
		\label{tab}
	\end{table}
	
	The constraints on the electromagnetic operators coming from the SINDRUM II results are weaker, by about an order of magnitude, than those previously obtained\cite{KPS} using the results of the MEG experiment at PSI, which probed $\mu^+ \rightarrow e^+ \gamma$.\cite{MEG} The MEG II experiment is currently taking data\cite{MEG2} and would be expected to further strengthen those bounds, although the later-stage results of COMET and Mu2e would likely produce competitive bounds for these operators.
	
	The constraints on the four-point quark-lepton operators are the first reported\cite{datatables} for these SME coefficients, which are experimentally accessible at tree level only in the nuclear conversion channel; this channel thus remains complementary to the other $\mu-e$ conversion searches. These constraints are expected to be improved by an order of magnitude or more from just the initial phases of the upcoming COMET and Mu2e experiments.
	
	\section*{Acknowledgments}
	This contribution to the proceedings is based on Ref.\ \refcite{orig}, in collaboration with V.\ Alan Kosteleck\'{y}, Emilie Passemar, and Nathaniel Sherrill. This contribution is supported in part by the U.S.\ National Science Foundation under grant PHY-2013184.


\begin{thebibliography}{xx}
		
		\bibitem{br}
		W.J.\ Marciano and A.I.\ Sanda, Phys.\ Lett.\ B {\bf 67}, 303 (1977); S.M.\ Bilenky, S.T.\ Petcov, and B.\ Pontecorvo, Phys.\ Lett.\ B {\bf 67}, 309 (1977); B.W.\ Lee, S.\ Pakvasa, R.E.\ Schrock, and H.\ Sugawara, Phys.\ Rev.\ Lett.\ {\bf 38}, 937 (1977).
		
		\bibitem{sind}
		A.\ van der Schaaf, J.\ Phys.\ G: Nucl.\ Part.\ Phys.\ {\bf 29}, 1503 (2003).
		
		\bibitem{com}
		COMET Collaboration, R.\ Abramishvili \etal, PTEP {\bf 2020}, 033C01 (2020).
		
		\bibitem{mu2e}
		Mu2e Collaboration, K.\ Byrum \etal, Universe {\bf 2023}, {\it 9}, 54 (2023); R.H.\ Bernstein, Front.\ Phys.\ {\bf 27}, 1 (2019).
		
		\bibitem{chrg-dist}
		H.\ de Vries, C.W.\ de Jager, and C.\ de Vries, At.\ Data Nucl.\ Data Tables {\bf 36}, 495 (1987).
		
		\bibitem{nucl}
		R.\ Kitano, M.\ Koike, and Y.\ Okada, Phys.\ Rev.\ D {\bf 66}, 096002 (2002).
		
		\bibitem{SCF-1}
		R.\ Bluhm, V.A.\ Kosteleck\'y, C.D.\ Lane, and N.\ Russell, Phys.\ Rev.\ D {\bf 68}, 125008 (2003); Phys.\ Rev.\ Lett.\ {\bf 88}, 090801 (2002); V.A.\ Kosteleck\'y and M.\ Mewes, Phys.\ Rev.\ D {\bf 66}, 056005 (2002).
		
		\bibitem{KPS}
		V.A.\ Kosteleck\'y, E.\ Passemar, and N.\ Sherrill, Phys.\ Rev.\ D {\bf 106}, 076016 (2022).
		
		\bibitem{SCF-2}
		V.A.\ Kosteleck\'y, A.C.\ Melissinos, and M.\ Mewes, Phys.\ Lett.\ B {\bf 761}, 1 (2016); Y.\ Ding and V.A.\ Kosteleck\'y, Phys.\ Rev.\ D {\bf 94}, 056008 (2016).
		
		\bibitem{orig}
		V.A.\ Kosteleck\'y, W.P.\ McNulty, E.\ Passemar, and N.\ Sherrill, Phys.\ Lett.\ B \textbf{868}, 139755 (2025).
		
		\bibitem{MEG}
		MEG Collaboration, A.M.\ Baldini \etal, Eur.\ Phys.\ J.\ C {\bf 76}, 434 (2016).
		
		\bibitem{MEG2}
		MEG II Collaboration, K.\ Afanaciev \etal, Eur.\ Phys.\ J.\ C {\bf 84}, 216 (2024).
		
		\bibitem{datatables}
		{\it Data Tables for Lorentz and CPT Violation,}
		V.A.\ Kosteleck\'y and N.\ Russell,
		2025 edition,
		arXiv:0801.0287v18.
		
	\end{thebibliography}
\end{document}